\documentclass[a4paper,twoside]{article}

\usepackage{subfigure}
\usepackage{calc}
\usepackage{amssymb}
\usepackage{amstext}
\usepackage{amsmath}
\usepackage{multicol}
\usepackage{pslatex}
\usepackage{apalike}
\usepackage{fancyhdr}
\usepackage{scitepress}
\usepackage[small]{caption}
\usepackage{ifpdf}

\ifpdf
    \usepackage[pdftex,
        pdfpagemode={UseOutlines},
        pdftitle={SPECI-II},
        pdfauthor={Ilango Leonardo Sriram},
       pdfcreator={pdflatex},
        plainpages=false,
        colorlinks=false,
        bookmarks=true,
        bookmarksnumbered=true,
        pdfhighlight={/I},
        breaklinks=false,
        urlcolor={blue},
        citecolor={blue},
        linkcolor={blue}]{hyperref}
    \usepackage[pdftex]{graphicx}
\else
    \usepackage{graphicx}
    \DeclareGraphicsExtensions{.svg}
\fi

\usepackage[dvips]{epsfig}
\usepackage{pgf}
\graphicspath{{figures/}}
\usepackage{url}

\begin{document}
\onecolumn \maketitle \normalsize \vfill

\section{INTRODUCTION}
\label{sec:introduction}

\noindent We introduce Version 2 of SPECI (Simulation Program for Elastic Computing Infrastructure), a system for predictive simulation modelling of ultra-large-scale data-centres (DCs), i.e. warehouse-sized facilities containing hundreds of thousands of servers, as used to provide cloud computing services. 

The move toward cloud computing is driving the construction of ever bigger DCs. For example, Microsoft's latest cloud-computing DC in Chicago has an estimated budget of US\$500m and capacity for 224,000 blade-servers \cite{dc:size:miller}.
The scale of such facilities means that the designers of these facilities have to work with data from development and testing set-ups that are often several orders of magnitude smaller than the final product. But architectures and management policies that work on a few hundred servers may not scale well to facilities housing hundreds of thousands
\cite{dc:scalability:jogalekar}.
However, although predictive simulation models have become commonplace \cite{simulation:fourthParadigm},
there is no well-established simulator to evaluate DC designs. 
A realistic simulator is difficult to achieve, as it needs to accommodate many models, such as network connectivity or disk access models, even heterogeneity, but 
many of these models lack a
uniform definition: e.g. although many clouds use virtualization some use MapReduce.
We believe it will require a set of simulation tools each modelling aspects of the cloud, and present SPECI-2 for
modelling middleware policy distribution in virtualised cloud DCs.

This paper explaines the SPECI model, the changes over the previous version and the reasons for these changes, 
and details of the implementation.

%After , we summarize new results from the simulator, and discuss future work.
%Section \ref{sec:speci-1} gives information on the previous version of SPECI, and background information on the model for cloud DC. 
%Section \ref{sec:implementation} . 
%Section \ref{sec:replication} reproduces some of the results published from SPECI-1 to verify the functional compatibility, and Section \ref{sec:hierarchy} shows results of simulations making %use of the hierarchical data centre cabling.
%We finish this paper with a conclusion in Section \ref{sec:conclusion}.

\section{SPECI-2}
\label{sec:implementation}

SPECI-2's goal remains to answer the same questions and requirements brought to
SPECI-1 and described in \cite{sriram:Cloudcom2009}:
consistency in middleware policy distribution.
Among practitioners there is the understanding that middleware for ultra-large DCs can only operate
on a certain scale if it is broken into policies which are distributed to the managed components and executed locally, as opposed to the use of centralised control components. Because the middleware's settings and available resources change very frequently, and changes can originate at arbitrary locations,
new policies need to be continuously communicated to the nodes.
Core to this problem are \textit{communication }\textit{\textbf{protocols}},
which allow components in the DC to communicate to other components, and the
\textit{component-subscription network }\textit{\textbf{topology}}, 
where services follow status changes of a subset of other components, where dependencies exist, in form of \textit{subscriptions}. 

The SPECI simulation models a DC hosting a number \textit{n} of cloud services,
which are connected through the subscription network. Each of these services has
a state that can change at a rate \textit{f}. Based on the frequency of \textit{f}
and the update protocol in place, some services' subscriptions will become inconsistent
with the current state, and inconsistencies might be propagated through the network
before the system returns to a consistent state. Every unit time, SPECI provides
a monitoring probe of the current number of inconsistencies, and the number of
network packets dealt with by every component in the DC.

\subsection{Changes over SPECI-1}

SPECI-2 is the first public release revised from our experiences with our earlier experimental versions of SPECI, the source of several previous peer-reviewed publications \cite{sriram:Cloudcom2009,sriram:ICECCS2010,sriram:CompleNet}. 
In our initial work with SPECI we used a simplification of a single hop connectivity between components. Since then,
\cite{dc:barroso-hoelzle}
published details about the hierarchies in Google's DCs, and of
relations between network connection costs for interconnects.
A DC is now modelled to contain aisles or clusters; each aisle contains racks; each rack contains chasses; each chassis contains blades; and each blade contains or runs cloud services. The quantification of this hierarchy can be specified in the configuration of the simulation run. When a cloud service communicates with another cloud service, the communication now follows the component tree.

Second, SPECI-1 used to poll a Boolean status of aliveness, to see whether a subscribed
component was alive or had failed. This was a simplification to the polling of
policies that could be changed by any service, where the simulator measures the
consistency of middleware policies in place.
SPECI-2 now has an Integer representing the version number
of the current policy in place, and this version number can be incremented over
time. All nodes subscribed will need to know this update, as it could potentially
mean new security settings, or other new behaviour. To continue to accommodate
component failure, the version number 0 represents a failed service. Cloud Services
with policy version 0 thus no longer participate in updates and no longer generate any load.

A further new requirement for SPECI-2 is a non-functional requirement: SPECI-1
suffered from weak performance and in particular of a heavy memory footprint. 
At runtime, this type of simulation depends more on the system memory requirements more than on the available CPU cycles, as it is designed to run in-memory and even in current HPC centres it is not common to find nodes with more than 10gb per core.
For this reason, SPECI-2 no longer maintains a java object for every component and
every network link, but only for the cloud services which represent the components
at the leaf of the DC hierarchy tree, and no longer uses a generic DES simulation
engine with heavy weight multi-purpose queues, but uses a customised event queue
that is more efficient, because it uses knowledge of the character of the set-up
towards resource optimisation: only the nearest events in time are in sorted order.
This saves memory and computational requirements for the queue, as in the simulation
set-up most insert operations join the unsorted queue. Further, the one to one
relationship between components in the DC and java objects was removed
and aggregated in singleton classes containing the behavioural logic. To save further
memory, monitoring data is no longer kept in memory over run-time, but stored in
persistent files and analysed in post-processing.
The SPECI-2 simulator shows a JVM memory footprint of 5.5GB RAM when modelling a DC with $10^6$ Cloud-Services with 316 subscriptions each. This is remarkably smaller footprint than that of SPECI-1 which required 25GB for this configuration.

\subsection{Simulator Usage}

A typical simulation run involves three scripts. The first generates a set of properties files, one for each combination of configuration parameters that shall be simulated.
The SPECI-2 simulator takes such a configuration file as input, runs the simulation, and writes the monitoring probes to a comma separated values file.
The output of these runs are the monitoring probes, which output the
current simulation time, the number of consistent and of inconsistent services, the load, and the maximum local loads.
For easy portability, the output is written to files and not to a database.
Hence, for the post processing after the simulation runs,
the third script is used to merge the content of the many output files into one, analyse the data statistically, calculate means and confidence intervals, and finally to generate graphs. There are a few python scripts that create the graphs using matplotlib, the one to choose depends on the desired output.

\subsection{Implementation and Design Details}

The SPECI-2 java simulator is started with an argument that passes on the location of a configuration
file. The entry class is SimulationRunner. This class first reads the configuration file and sets the configuration parameters in a static class.
It then creates the utility objects required, e.g. those used for relevant random
draws to wire the subscriptions.
It then creates a structure object that contains the DC setup, hence it generates
the layout and components for the DC based on the configuration file.
This includes arrays for every type of physical objects, with the elements of the array keeping track of the load in form of access counts generated by every individual component of that type.
There are arrays containing elements for every aisle, and likewise for components at rack level, chassis level, and blade level.
For the Cloud Service
level components modelled, the structure object creates an object for for every Cloud Service, which
holds both integers for monitored
load as well as a pointer to the object of the relevant service.

There are two utility classes with public static methods, SubscriptionGenerator
and Protocol. Only the Persistence class contains both static methods and variables, and Configuration
has static variables that are read from file once initialised, this reduces the amount
of file access required. Finally, there is one singleton class that stores the
arrays and access counts. To continue the initialisation phase, the SimulationRunner entry class then calls a utility function that wires the subscriptions to each of the Cloud services
depending on the current configuration parameters and then initialises the queue.
Once the datacentre is initialised, the execution of the simulation is entirely driven by the queue.
After the execution of an event it will schedule itself for its next update.

The simulation queue is a custom queue which holds tuples of
time and int, with positive integers referring to the id of a cloud service and
negative integers being predefined events other than updates, such as events for
changes to occur, or monitoring probes to being taken.
For performance reason,
the queue is divided into two array lists:
a sorted list for those events to be executed
shortly, which always has the next event in time at the beginning, and an unsorted list for events further away.
This promises performance advantages,
as the nature of the experiment is such that most of the newly arriving events
will be further away in time than the time of the mean of the events in the queue:
thus for most insert operations costly sorted inserts can be avoided.

In summary, SPECI-2 gained performance, readability and extensibility, at the cost of style: 
some components have centralised knowledge although the simulator models a
decentralised DC. The use of singleton helps performance, but on the other
it makes integration testing very difficult.  

\section{REPLICATION OF SPECI-1}
\label{sec:replication}

\noindent To confirm that SPECI-2's outcome is in line with SPECI-1, we have constructed experiments mimicking the flat model used in SPECI-1.
To achieve such a setup without any hierarchies and providing a one-hop connectivity, in this section we model all cloud services of the entire DC to fit on a single blade, and set the unit cost of communicating to another cloud service on the same blade to be 1 per access count. This way we could reproduces some of the results
published from SPECI-1 to verify the simulator to be compatible with SPECI-1.

The model observes the number of nodes that have an inconsistent view of the system. A node has an inconsistent view if any of the subscriptions that node has contains incorrect aliveness information. The number of inconsistent nodes is              measured over time and observed once every $\Delta$t (=1sec).

For the graphs shown here, we assume that the number of subscriptions grows slower than the total size of the DC, and so we set the average number of subscriptions per node to $\sqrt{n}$.
For each of these sizes a failure or change rate distribution
\textit{f} was chosen such that on average over the runtime 0.01\%, 0.1\%, 1\%, and 10\%
of the nodes would fail. The graphs contain the half-width of the 95\% confidence intervals, which for the load graphs however are small and barely visible in the graph.

Figure \ref{initSpeciProtocolsInc} and Figure \ref{initSpeciProtocolsLoad} 
show the effect of the subscription graph topology on the levels of inconsistencies, see \cite{sriram:ICECCS2010} 
for an explanation of the topology networks and the original figures from SPECI-1.
If the subscriptions graph has the structure of a Random or Barabasi-Albert graph, the distribution is more resilient towards transitive passing-on of inconsistencies than with Strogratz-Watts or Regular graphs. On the other hand it also generates a significantly higher load. This shows that the nature of the subscription graph, which is intrinsic by the jobs that reside on the DC, needs to be taken into account when tuning the middleware.
Similar replication has been made to reconfirm other graphs previously published but will not be shown here.

\section{HIERARCHICAL DC}
\label{sec:hierarchy}

\noindent This section shows further results of simulations of the previous scenario on a hierarchically wired DC. As an exploratory hierarchy we compare the previous results with a DC set-up of 4 cloud services per blade, 16 blades per chassis, 4 chasses per rack and 16 racks per aisle, and we leave it to the future work to investigate the effect of varying these hierarchies.

\begin{figure}[pt]
\begin{center}
\includegraphics[width=0.4\textwidth, height=30mm]{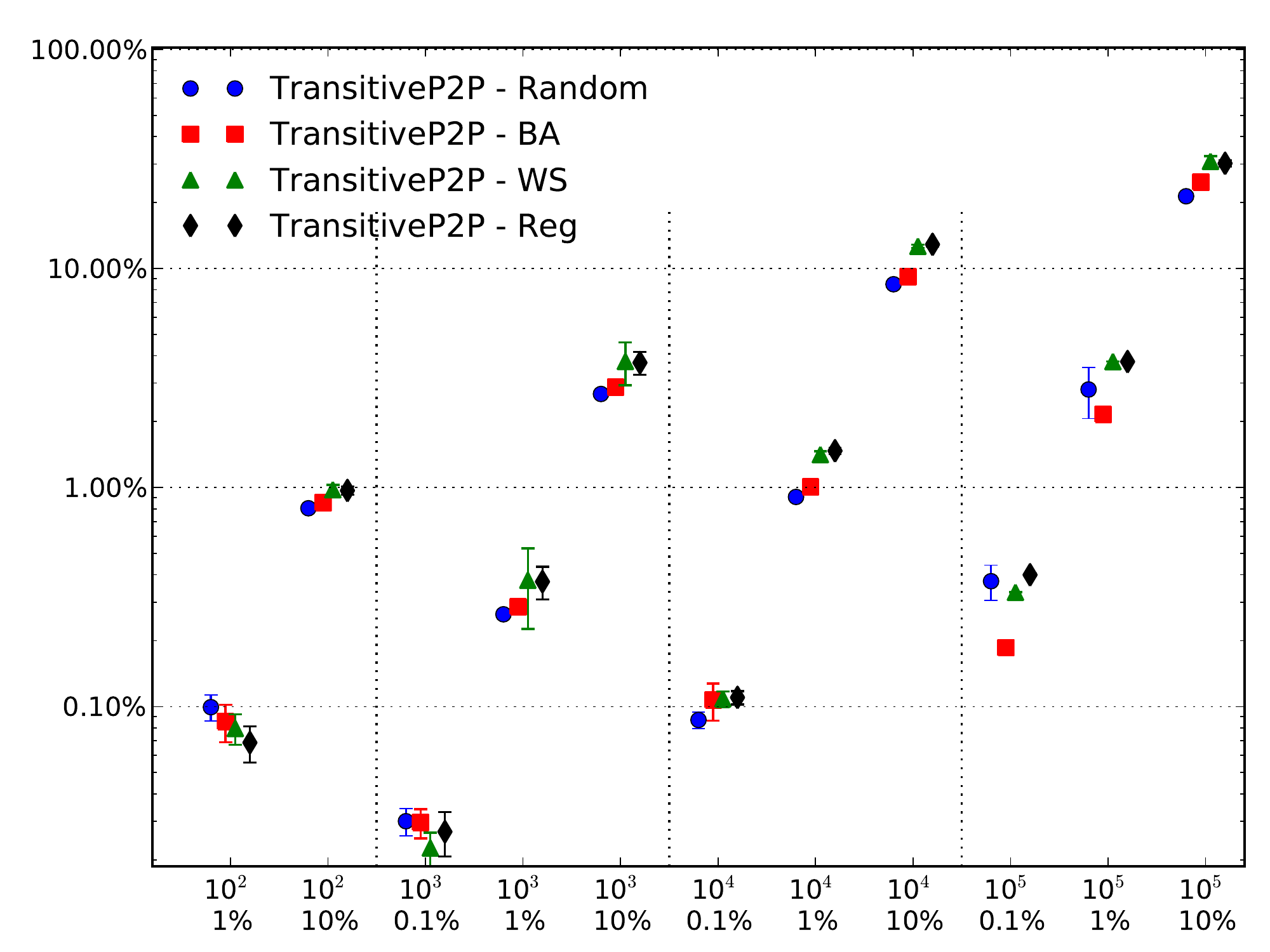}
\caption{Depending on the nature of the subscription graph, the middleware exhibits variations in the number of inconsistencies.}
\label{initSpeciProtocolsInc}
\end{center}
\end{figure}

\begin{figure}[pt]
\vspace{-20pt}
\begin{center}
\includegraphics[width=0.4\textwidth, height=30mm]{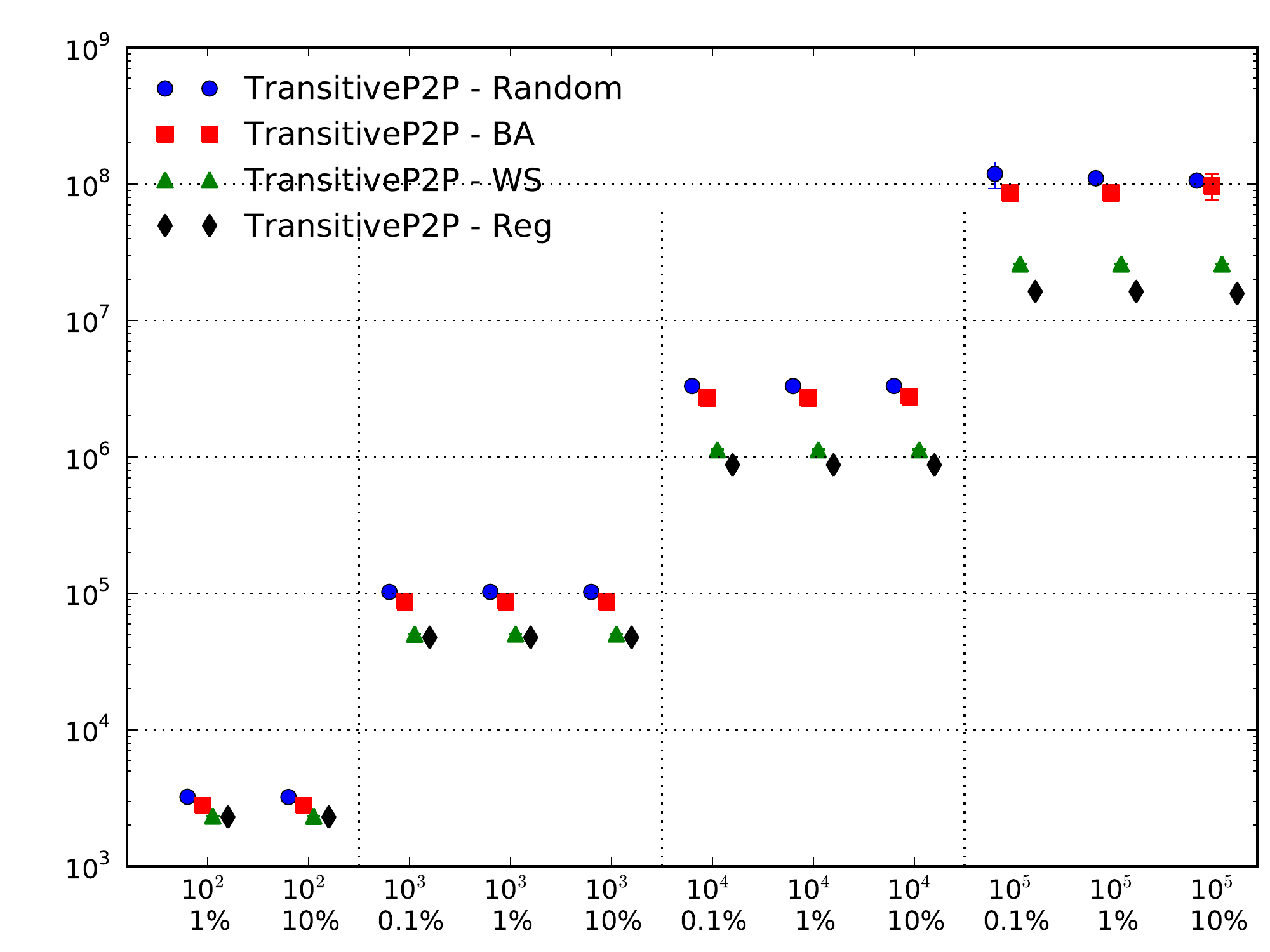}
\caption{If the distribution of the subscriptions is of a regular graph or Strogratz-Watts network, they require a higher load, which offsets their advantages in terms of inconsistencies}
\label{initSpeciProtocolsLoad}
\end{center}
\end{figure}

\begin{figure}[pt]
\begin{center}
\includegraphics[width=0.4\textwidth, height=30mm]{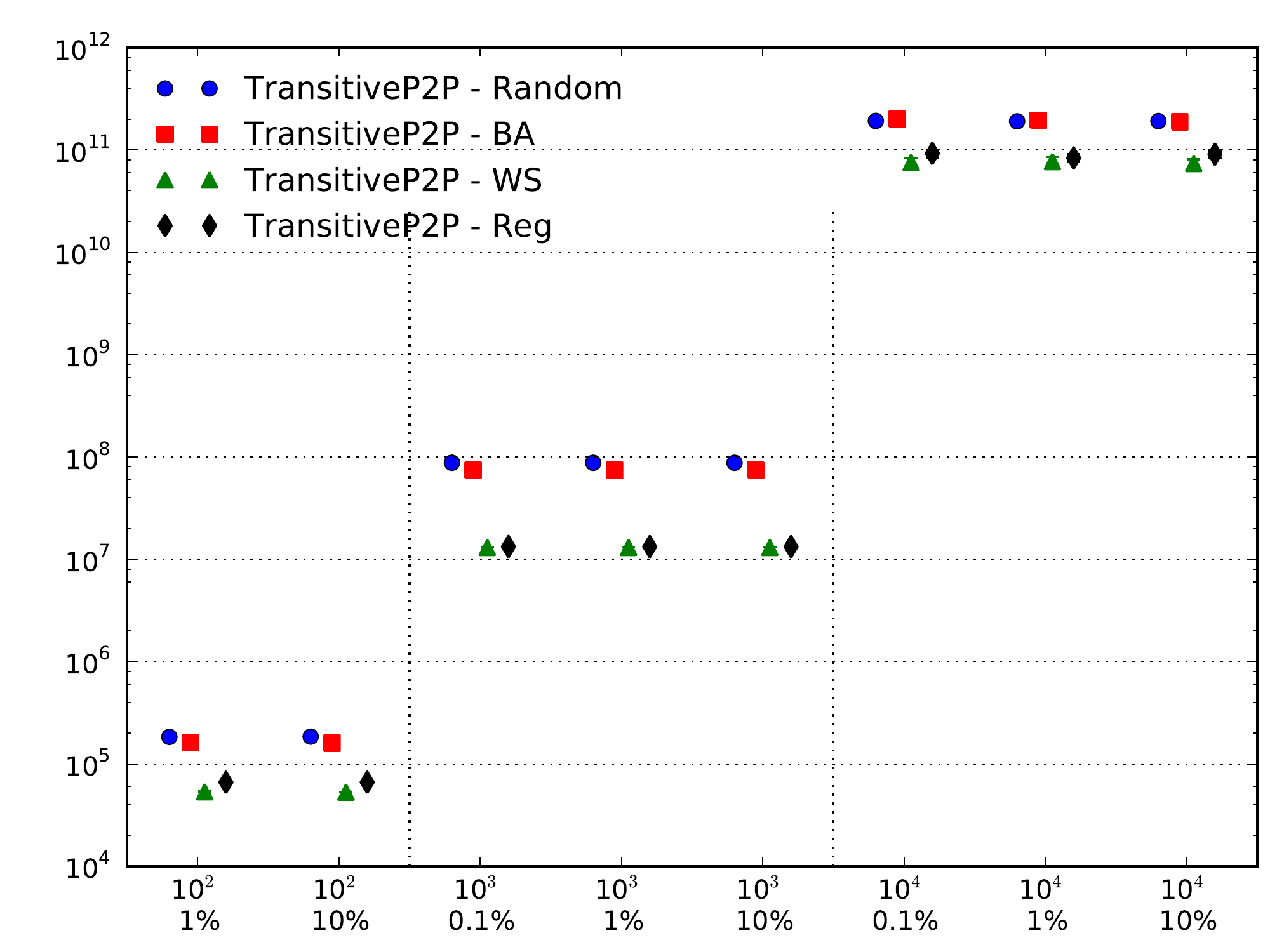}
\caption{The load on the hierarchical DC is much higher than the load in Figure \ref{initSpeciProtocolsLoad}, as potentially multiple hops are required for every communication, and as costs are introduced. This feature allows us to model communication and management cost of placement strategies. }
\label{speciH4Load}
\end{center}
\end{figure}

The consistency graph of the TransitiveP2P protocol for the various subscription topologies is essentially identical to the one in 
Figure \ref{initSpeciProtocolsInc}, which shows a ``flat'' DC. This is due to the fact that the logical layer that deals with the
communication protocol is not affected by the physical layout, as it still continues to communicate with the same other services as in the flat scenario. For this experiment, the placement and choice of subscriptions is dependent on the network subscription topology graphs, and is not correlated with the geographical distance in the DC.
On the other hand Figure \ref{speciH4Load} shows a much higher load count than Figure \ref{initSpeciProtocolsLoad}, as potentially multiple hops are required for every communication, and as costs are introduced. Note, compared to the experiments reported in \cite{sriram:ICECCS2010} here only those services that do not experience failure over the simulation time are counted towards the average load and average inconsistencies. This makes the load entirely independent of the change rate. In Figure \ref{speciH4Load}
one can observe that unlike in the flat DC the load does not increase by a constant factor. The step from $10^3$ to $10^4$ is more than an order of magnitude bigger than the step from $10^2$ to $10^3$. This is a direct effect of the scale requiring longer communication paths, and more subscriptions being further away. This type of observation can allow us to model communication and management cost of placement strategies in the future work.

\section{CONCLUSIONS}
\label{sec:conclusion}

In this paper we have introduced SPECI-2. It has benefits in performance and extensibility, and it models a hierarchical DC.
We have demonstrated both the need for such tools as well as a simulation architecture suitable for a hierarchical DC layout.
With the release of SPECI-2 we are hoping to attract a community of researchers interested in modelling aspects of DCs.
We have further shown that SPECI-2 is compatible with the results published using SPECI-1 and have shown areas of investigation that can be followed with SPECI-2 in the future.

%\section*{\uppercase{Acknowledgements}}
%\noindent This project is part of a PhD funded by Hewlett-Packard Labs' Automated Infrastructure
%Lab. We thank Hewlett-Packard for the interest in this topic and support.

\renewcommand{\baselinestretch}{0.98}
\bibliographystyle{apalike}
{\small
\bibliography{bibliography}}

\begin{thebibliography}{}

\bibitem[Barroso and H\"{o}lzle, 2009]{dc:barroso-hoelzle}
Barroso, L.~A. and H\"{o}lzle, U. (2009).
\newblock The datacenter as a computer: An introduction to the design of
  warehouse-scale machines.
\newblock {\em Synthesis Lectures on Computer Architecture}, 4(1):1--108.

\bibitem[Hey et~al., 2009]{simulation:fourthParadigm}
Hey, T., Tansley, S., and Tolle, K. (2009).
\newblock {\em The Fourth Paradigm: Data-Intensive Scientific Discovery}.
\newblock Microsoft.

\bibitem[Jogalekar and Woodside, 2000]{dc:scalability:jogalekar}
Jogalekar, P. and Woodside, M. (2000).
\newblock Evaluating the scalability of distributed systems.
\newblock {\em IEEE Transactions on Parallel and Distributed Systems},
  11(6)(11:6):589--603.

\bibitem[Miller, 2009]{dc:size:miller}
Miller, R. (2009).
\newblock Microsoft unveils its container-powered cloud.
\newblock
  \url{http://www.datacenterknowledge.com/archives/2009/09/30/microsoft-unveil%
s-its-container-powered-cloud/}.

\bibitem[Sriram, 2009]{sriram:Cloudcom2009}
Sriram, I. (2009).
\newblock Speci, a simulation tool exploring cloud-scale data centres.
\newblock In {\em Cloud Computing: First International Conference, CloudCom
  2009, LNCS 5931}, pages 381--392, Beijing, China. Springer-Verlag Berlin
  Heidelberg.

\bibitem[Sriram and Cliff, 2010]{sriram:ICECCS2010}
Sriram, I. and Cliff, D. (2010).
\newblock Effects of component-subscription network topology on large-scale
  data centre performance scaling.
\newblock In {\em Proceedings of the 15th IEEE International Conference on
  Engineering of Complex Computer Systems (ICECCS 2010)}, pages 72 -- 81,
  Oxford, UK. IEEE Computer Society.

\bibitem[Sriram and Cliff, 2011]{sriram:CompleNet}
Sriram, I. and Cliff, D. (2011).
\newblock Hybrid complex network topologies are preferred for
  component-subscription in large-scale data-centres.
\newblock In da~F. Costa~et al., L., editor, {\em CompleNet 2010}, volume 116
  of {\em Communications in Computer and Information Science}, pages 130--137.
  Springer.

\end{thebibliography}
\renewcommand{\baselinestretch}{1}

\end{document}